\documentclass[aip,
 amsmath,amssymb,
 reprint,%
]{revtex4-1}

\usepackage{graphicx}
\usepackage{dcolumn}
\usepackage{bm}

\usepackage[utf8]{inputenc}
\usepackage[T1]{fontenc}
\usepackage{mathptmx}
\usepackage{etoolbox}

\makeatletter
\def\@email#1#2{%
 \endgroup
 \patchcmd{\titleblock@produce}
  {\frontmatter@RRAPformat}
  {\frontmatter@RRAPformat{\produce@RRAP{*#1\href{mailto:#2}{#2}}}\frontmatter@RRAPformat}
  {}{}
}%
\makeatother
\begin{document}

\preprint{AIP/123-QED}

\title[]{Nanoscale magnetic field sensing with spin-Hall nano-oscillator devices}
\author{Yanyou Xie}
\affiliation{%
School of Applied \& Engineering Physics, Cornell University, Ithaca, New York 14850, USA
}%
\author{Hil Fung Harry Cheung}%
\affiliation{ 
Department of Physics, Cornell University, Ithaca, New York 14850, USA
}%

\author{Gregory D. Fuchs}
\email{gdf9@cornell.edu.}
\affiliation{%
School of Applied \& Engineering Physics, Cornell University, Ithaca, New York 14850, USA
}%

\date{\today}

\begin{abstract}
In this work, we develop a nanoscale magnetic field sensor based on a spin-Hall nano-oscillator (SHNO) device.   We fabricate constriction-based SHNO devices using Ni$_{81}$Fe$_{19}$/Au$_{0.25}$Pt$_{0.75}$ bilayer and use them to demonstrate quasi-DC sensing up to kilohertz-scale frequencies under a bias field of 400 Oe in the sample plane. The magnetic field sensing is based on the linear dependence of SHNO oscillation frequency on external magnetic field. 
The detectivity of the sensor is as low as 0.21 $\mu$T/$\sqrt{\mbox{Hz}}$ at 100 Hz, with an effective sensing area of 0.32 $\mu$m$^2$, making SHNO devices suitable for scanning probe nanoscale magnetometry.
\end{abstract}

\maketitle

Nanoscale magnetic field sensing is essential for many applications, including high-density magnetic storage readout, biomagnetic signal detection, and high resolution magnetic imaging.\cite{braganca2010nanoscale,cao2020development,hong2013nanoscale} 
Over the years, a range of sensors and sensor-based magnetometry techniques have been developed using distinct physical principles, such as magnetoresistance (MR),\cite{reig2013giant,weiss2013advanced,cao2020development} the Hall-effect,\cite{ramsden2011hall,chang1992scanning,sandhu2004nano} superconducting quantum interference devices,\cite{kirtley1995high,kirtley1999scanning,weinstock2012squid} and diamond nitrogen-vacancy (NV) centers.\cite{hong2013nanoscale,degen2008scanning} 
In addition to the aforementioned sensors, the oscillations of nanoscale magnetization driven by the spin-transfer torque \cite{ralph2008spin} from spin-polarized current in an electrical device (a spin-torque nano-oscillator, STNO) have been proposed as a nanoscale sensing candidate. \cite{braganca2010nanoscale, albertsson2020magnetic, chen2014inductorless, chen2016spin} However, the quantification of the sensor detectivity in such devices is lacking.

Recently, a new class of magnetic nano-oscillator --- spin-Hall nano-oscillators (SHNOs)\cite{liu2012magnetic,demidov2012magnetic,demidov2014nanoconstriction,duan2014nanowire,zhang2021spin} --- have been developed with an open geometry that can be straightforwardly scaled into arrays and networks.\cite{awad2017long,zahedinejad2020two}  As a result, they have proven promising for the emerging field of neuromorphic computing.\cite{zahedinejad2020two,zahedinejad2022memristive} Similar to STNOs, SHNOs use spin torque to fully compensate the intrinsic damping in magnetic materials, leading to magnetic auto-oscillations in the GHz frequency regime. \cite{demidov2012magnetic} The spin torque in SHNOs originates from the spin Hall effect, where a dc charge current in a nonmagnetic (NM) layer with spin-orbit interaction generates a pure spin current along transverse directions that flows into an adjacent ferromagnetic (FM) layer.\cite{hirsch1999spin,sinova2015spin} SHNOs of different designs  \cite{demidov2012magnetic,liu2012magnetic,demidov2014nanoconstriction,duan2014nanowire,zhang2021spin,sato2019domain} have been developed. Out of these designs, we focus on constriction-based SHNOs\cite{demidov2014nanoconstriction} because they have a simple fabrication process that enables 1-D or 2-D arrays. \cite{awad2017long,awad2020width,zahedinejad2020two, kendziorczyk2016mutual, mazraati2016low,divinskiy2017nanoconstriction} SHNOs respond to external stimuli on a nanosecond time scale, \cite{gonccalves2021agility} and have a tunable oscillation frequency as a function of dc charge current  \cite{zahedinejad2018cmos,divinskiy2017nanoconstriction} and external magnetic field.\cite{dvornik2018origin} With a local linear response of oscillation frequency with respect to external magnetic field, \cite{dvornik2018origin} SHNOs provides a solid basis for quasi-DC magnetic field sensing. By monitoring the oscillation frequency, we measure variations in the local magnetic field.

In this work, we study the temporal stability of constriction-based SHNOs and demonstrate the implementation of magnetic field sensors based on these devices. We characterize quasi-DC sensing up to kilohertz-scale frequencies. The magnetic field sensor has a detectivity of $0.21\ \mu\mbox{T}/\sqrt{\mbox{Hz}}$ at 100 Hz, with an effective sensing area of $0.32\ \mu\mbox{m}^2$. The devices are able to sense small magnetic field variations around a 400 Oe bias field. The nanoscale sensing area of the SHNO devices makes them interesting as local sensors, for example in scanning probe magnetometry. 

\begin{figure}
\includegraphics[width=0.95\linewidth]{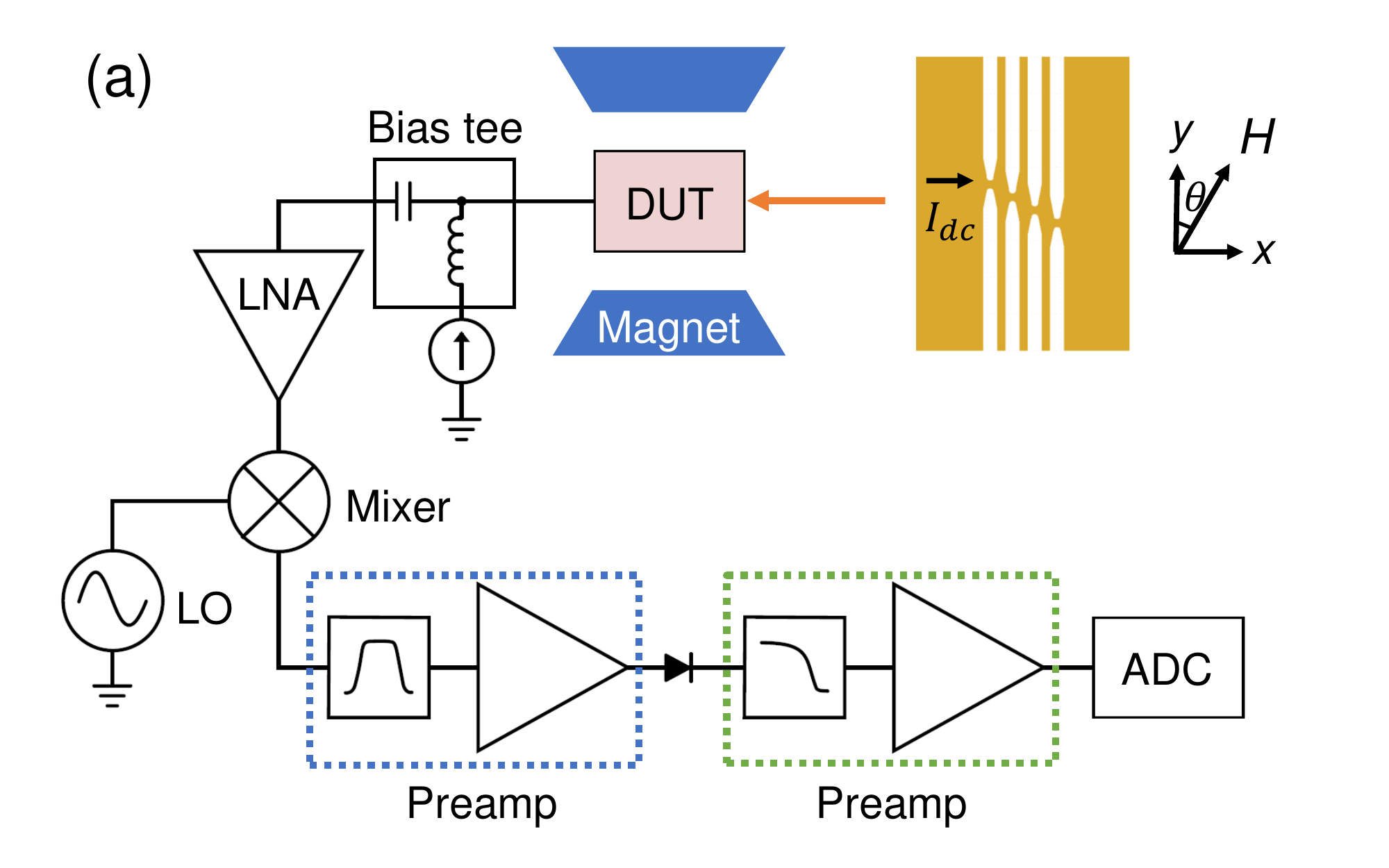}\\
\includegraphics[width=\linewidth]{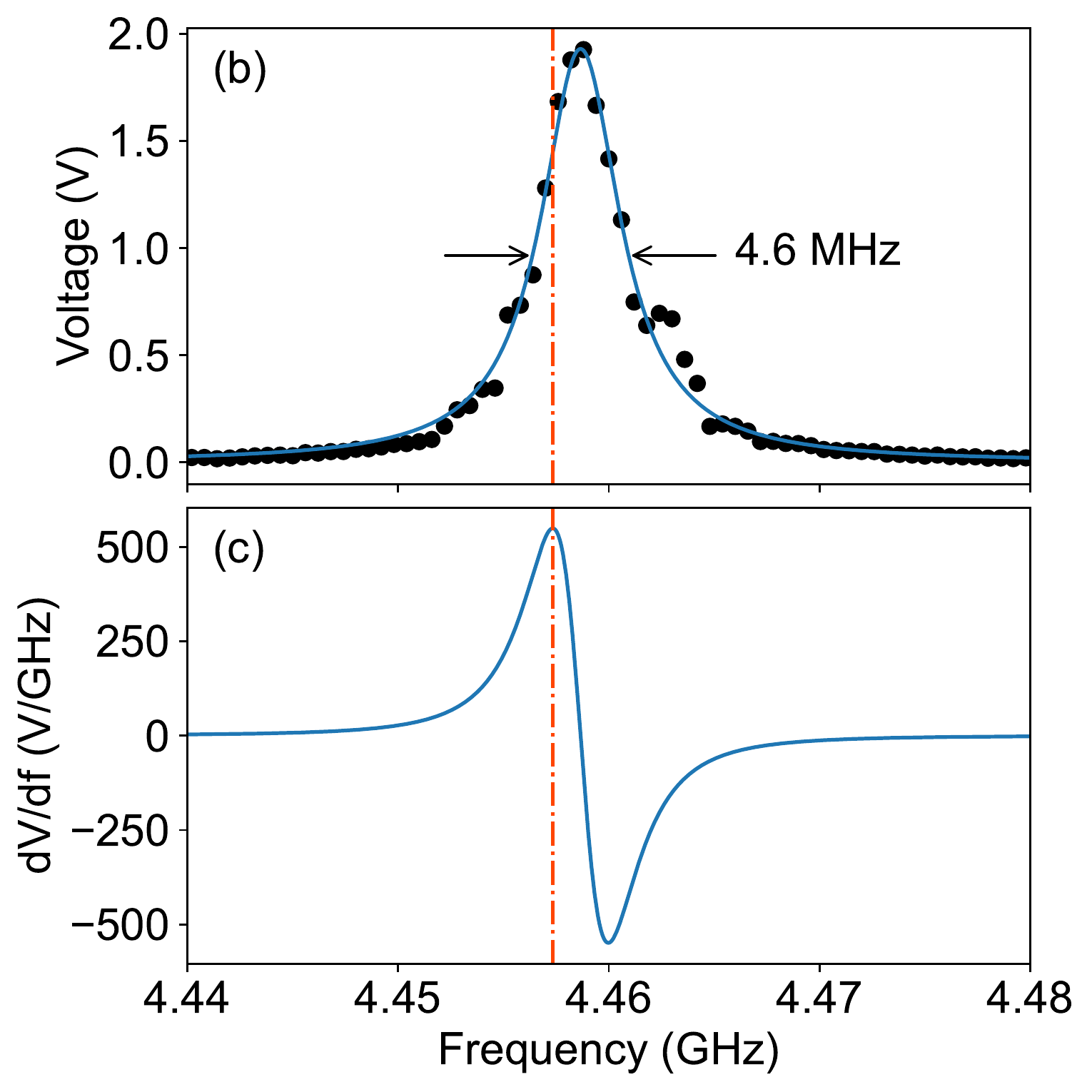}
\caption{\label{Circuit} (a) Schematic of the experiment with device geometry. (b) A sample frequency scan of 4-constriction device at $H=400$ Oe and $I_{dc}=1.67$ mA with Lorentzian fit, showing a FWHM of 4.6 MHz. (c) The derivative of the Lorentzian fit in (b), with the frequency of the maximum slope shown in dotted line}
\end{figure}

Our sensors include both 1- and 4-constriction devices. The 150-nm-wide nanoconstrictions are patterned in a bow-tie shape, similar to previous studies.\cite{demidov2014nanoconstriction, kendziorczyk2016mutual, mazraati2016low, awad2017long, divinskiy2017nanoconstriction} The 4-constriction device [Fig.~\ref{Circuit}(a)] is designed to promote mutual synchronization, with the constrictions aligning at an angle perpendicular to the external magnetic field, which maximizes spin wave overlap.\cite{awad2017long, kendziorczyk2016mutual} The nanoconstrictions in the 4-constriction device are separated by 350 nm center-to-center.

The devices are based on a Ni$_{81}$Fe$_{19}$(5)/Au$_{0.25}$Pt$_{0.75}$(5) bilayer (thicknesses in nanometers) sputtered on a sapphire substrate. We choose the heavy metal alloy Au$_{0.25}$Pt$_{0.75}$ because of its high spin Hall conductivity. \cite{zhu2018highly} We pattern the bilayer into $20.5\ \mu\mbox{m}\times 4 \ \mu\mbox{m}$ bars with 1 or 4 nanoconstrictions using e-beam lithography and Ar ion milling. The shape of electrical contact pads are then defined by a contact aligner, followed by thermal evaporation of Cr/Au and liftoff. 

Electrical characterization and magnetic field sensing measurements are made using the home-built spectrum analyzer setup shown in Fig.~\ref{Circuit}(a). This setup is capable of recovering the output power of SHNOs, which is in the pW/GHz level. The gigahertz  frequency auto-oscillation of magnetization is converted to a gigahertz electric signal through the anisotropic magnetoresistance (AMR) effect. \cite{demidov2014nanoconstriction} Because the AMR signal size is dependent on the cosine squared of the angle between the magnetization and the charge current direction, we set the external magnetic field $H$ to be $30^\circ$ from the $y$ axis. A dc charge current $I_{dc}$ is applied to the device to excite auto-oscillations. The gigahertz output voltage is amplified and down-converted to a megahertz frequency voltage by mixing it with a local oscillator (LO). The resulting signal then goes through a band pass filter of 1 Hz-300 kHz, which defines a resolution bandwidth (RBW) of 600 kHz. After the band pass filter, the signal is amplified and converted to a voltage signal by an RF diode. Then the voltage signal goes through a 3 kHz low pass filter, which defines the video bandwidth (VBW). The voltage signal is amplified again before being digitized.

To spectrally resolve auto-oscillations, we sweep the frequency of the LO and record the corresponding output signal from the SHNO device. When the frequency of the SHNO device matches the frequency of LO within the RBW, a peak appears, as shown in Fig.~\ref{Circuit}(b). We fit the peak to a Lorentzian to extract its linewidth. The derivative of the Lorentzian fit [Fig.~\ref{Circuit}(c)] gives information on the relationship between the final voltage collected and oscillation frequency.

\begin{figure}
\includegraphics[width=\linewidth]{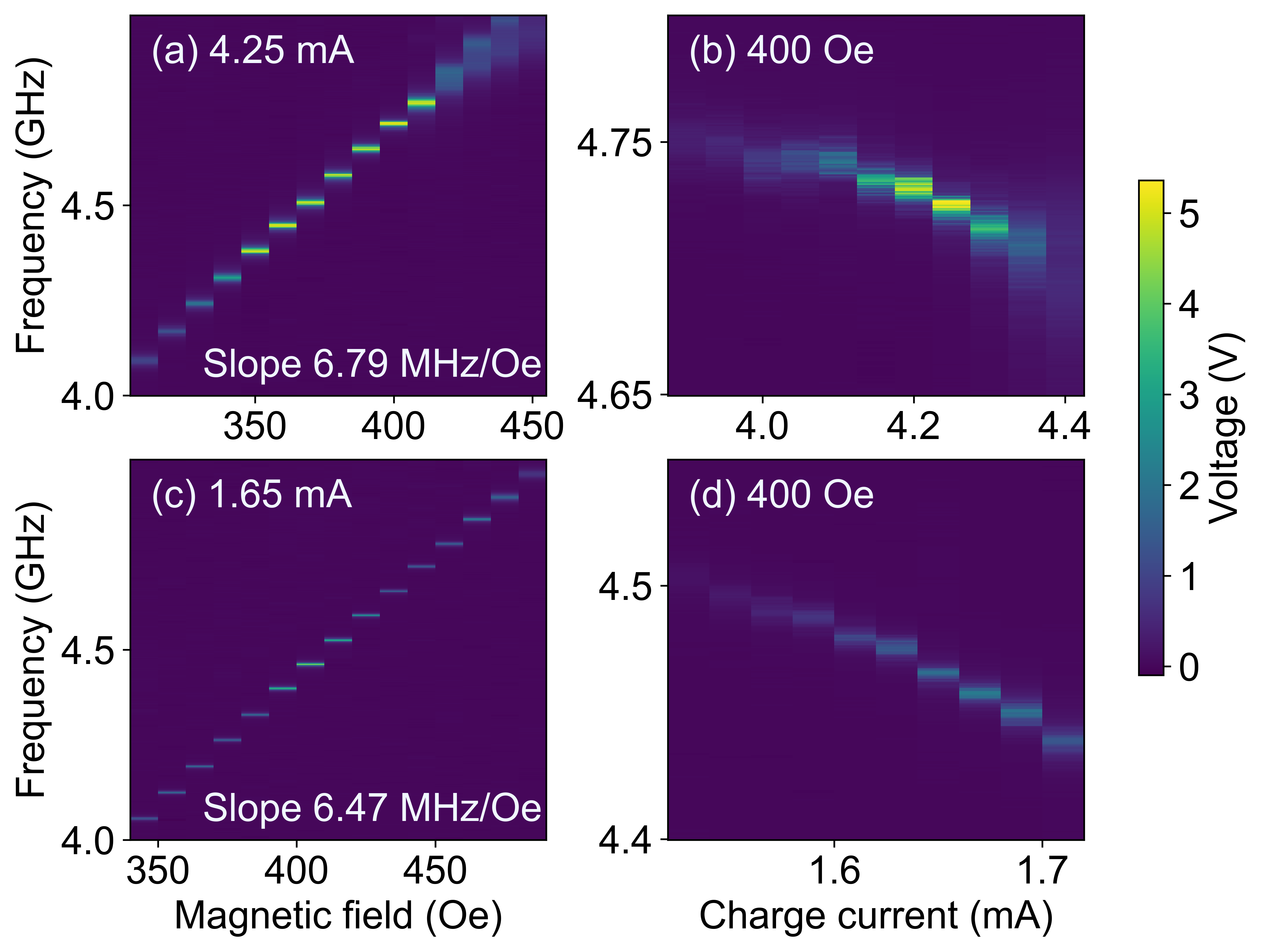}
\caption{\label{Auto} Auto-oscillation frequency dependence on (a) external magnetic field at $I_{dc}=4.25$ mA and (b) charge current at $H=400$ Oe for 1-constriction device. (c)(d) shows the frequency dependence of external magnetic field and charge current for 4-constriction device.}
\end{figure}

We acquire auto-oscillation spectra from a device with 1 constriction [Fig.~\ref{Auto}(a)(b)] and a device with 4 constrictions [Fig.~\ref{Auto}(c)(d)], respectively. The 1-constriction device has a resistance of 143 $\Omega$ and the 4-constriction device has a resistance of 451 $\Omega$. 
As shown in Fig.~\ref{Auto}, the power output of the peak depends on $I_{dc}$ and $H$. We identify the optimal $I_{dc}$ and $H$ combination where each SHNO devices has its maximum output power. For the 1-constriction device that is $I_{dc}=4.25$ mA and $H = 400$ Oe, and for the 4-constriction device it is $I_{dc}=1.67$ mA and $H=400$ Oe. The 1-constriction device has a larger output power than the 4-constriction device, mainly due to its better impedance match to the 50 $\Omega$ transmission line. Both devices show a positive linear dependence of oscillation frequency on external magnetic field [Fig.~\ref{Auto}(a)(c)], with similar slopes --- $\partial f/\partial H= 6.79$  MHz/Oe for the 1-constriction device and $\partial f/\partial H = 6.47$ MHz/Oe for the 4-constriction device.  
This linear dependence enables the application of SHNO devices as magnetic field sensors. Upon closer examination, we notice that the 4-constriction device has a smaller linewidth over the magnetic field range that we study. In contrast, the linewidth of the 1-constriction device significantly broadens above 420 Oe. The 4-constriction device may have synchronization,\cite{awad2017long} which would lead to increased tolerance to phase noise and thus lead to smaller linewidth over a larger magnetic field range. Although the oscillation frequencies of the two devices are slightly different, they exhibit a similar decrease in the oscillation frequency with an increase in $I_{dc}$ [Fig.~\ref{Auto}(b)(d)], consistent with other studies \cite{demidov2014nanoconstriction,kendziorczyk2016mutual,mazraati2018auto,gonccalves2021agility,zhang2021spin} of SHNO devices under in-plane magnetic field. This frequency red-shift is due to the negative magnetodynamic nonlinearity for small amplitude oscillations under in-plane magnetic field. \cite{slavin2009nonlinear,zhang2021spin,awad2020width}

\begin{figure*}
\includegraphics[width=\linewidth]{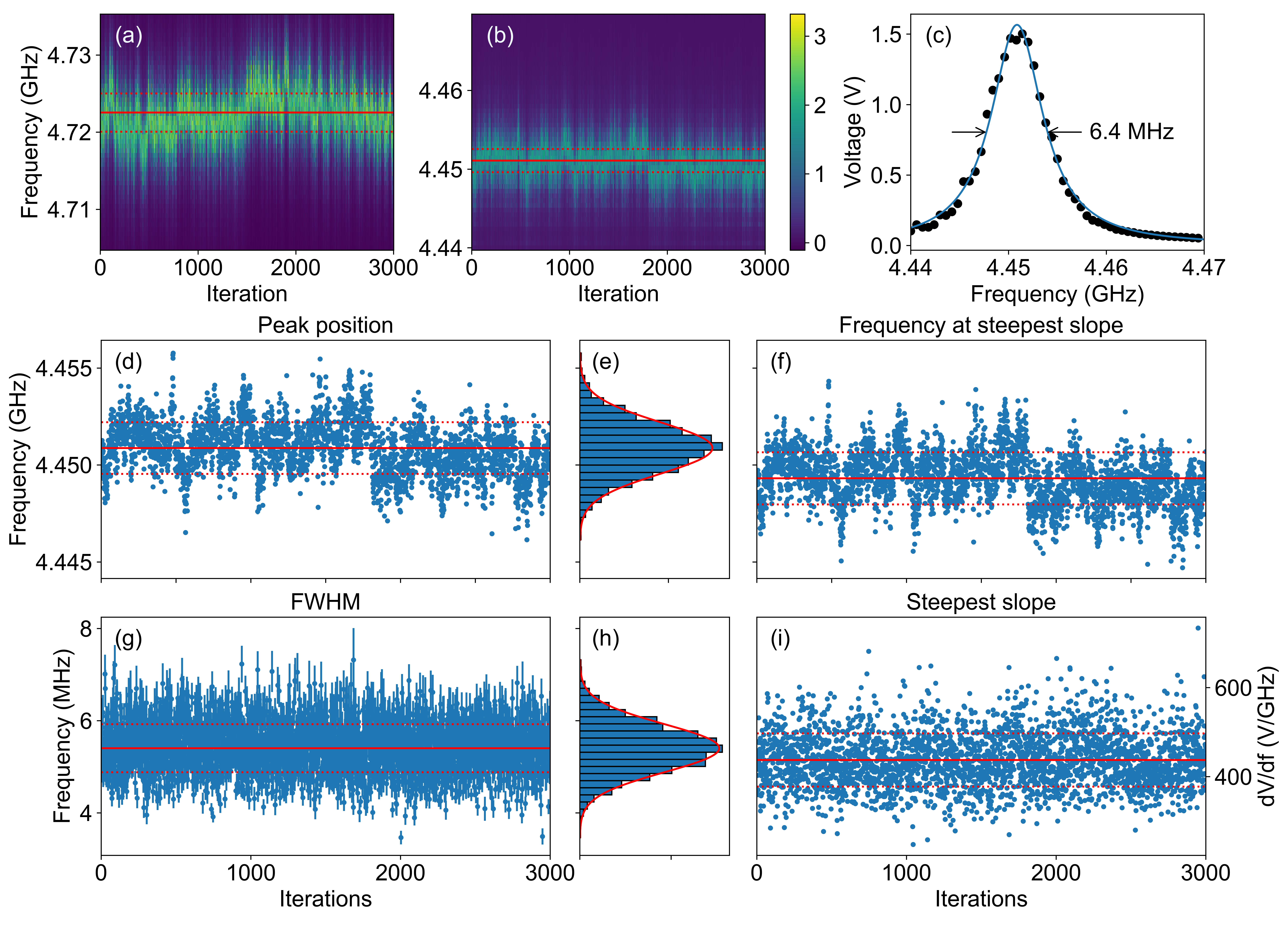}
\caption{\label{3000} Frequency scan over 3000 iterations for (a) 1-constriction and (b) 4-constriction device. (c) Average signal over 3000 iterations with Lorentzian fit for 4-constriction device. Individual scans in (b) are fitted by Lorentzian function and extracted (d) peak position, (f) frequency at steepest slope, (g) FWHM and (i) steepest slope. The solid lines represent the means and the dotted lines represent 1 standard deviation from the means. (e) and (h) are histograms of (d) and (g), respectively. The solid lines are Gaussian distribution with the sample mean and variance. }
\end{figure*}

To use the SHNO devices as magnetic field sensor, we want maximum sensitivity. Sensitivity $S$ is the derivative of the output electrical signal with respect to the input physical signal,\cite{wilson2004sensor} here 
\begin{equation}
S = \frac{\partial V}{\partial H}=\frac{\partial V}{\partial f}\frac{\partial f}{\partial H},
\end{equation}
where $V$ is the output voltage signal, $f$ is the oscillation frequency of the SHNO device, $H$ is the external magnetic field. $\partial V/\partial f$ is the maximum value on the derivative of Lorentzian fit, such as in Fig.~\ref{Circuit}(c), while $\partial f/\partial H$ is the slope extracted from Fig.~\ref{Auto}(a)(c). It is likely that $\partial V/\partial f$ is larger for higher output power, whereas $\partial f/\partial H$ is fixed for a specific SHNO device. Therefore, we operate our SHNO devices at the optimal $I_{dc}$ and $H$ combination, where the devices are likely to have largest sensitivity. 

The principle of sensing is based on the change in oscillation frequency of SHNOs with magnetic field. If we fix the frequency of LO somewhere on the peak (such as in Fig.~\ref{Circuit}(b)), the output voltage changes when magnetic field varies around the 400 Oe bias field as a result of oscillator frequency shift. This way we can extract the variation of magnetic field by measuring the output voltage change. We choose the frequency of the LO to correspond to the steepest slope at the optimal $I_{dc}$ and $H$ combination, where the sensor has largest sensitivity. To find the required frequency, we measure the SHNO spectrum at the bias conditions and fit a Lorentzian.

After identifying the operating frequency, we characterize our SHNO sensors. We observe a spectral drift in the oscillation frequency when acquiring 3000 spectra over about 3 hours [Fig.~\ref{3000}(a)(b)]. Note that the effective sensing area is only 0.071 $\mu$m$^2$ for the 1-constriction device, and 0.32 $\mu$m$^2$ for the 4-constriction device. Because of their small auto-oscillation mode volume, the devices are susceptible to thermal fluctuations, which displace the precessing magnetization along and transverse to its steady-state orbit, leading to a spread in oscillation frequency.\cite{chen2016spin,sankey2005mechanisms} Externally, the devices also subject to fluctuation of the bias magnetic field, which also contributes to a spectral drift. 
The overall drift is more prominent in 1-constriction device [Fig.~\ref{3000}(a)] than 4-constriction device [Fig.~\ref{3000}(b)], consistent with the idea that a larger oscillator volume protects it from fluctuations. 

For sensing, it is desirable that the peak frequency is as stable as possible, so the 4-constriction device is a better candidate. Therefore, the following sensing measurements are focused on the 4-constriction device. We analyze the 3000 spectra of the 4-constriction device by fitting Lorentzian function to individual scans in Fig.~\ref{3000}(b) to extract their peak position [Fig.~\ref{3000}(d)] and full-width-half-maximum (FWHM) [Fig.~\ref{3000}(g)]. We also calculate the frequency at the steepest slope [Fig.~\ref{3000}(f)] and the corresponding slope magnitude [Fig.~\ref{3000}(i)]. We see fluctuations of these variables over different scans. The distribution of the peak position [Fig.~\ref{3000}(e)] and FWHM [Fig.~\ref{3000}(h)] are close to Gaussian, indicating the random nature of the frequency fluctuation. The frequency fluctuation is characteristic of a nonlinear auto-oscillator in which the oscillation frequency depends strongly on the amplitude.\cite{kim2008line} We also fit a Lorentzian to the average signal over 3000 iterations [Fig.~\ref{3000}(c)] and extract $\partial V/\partial f=316$ V/GHz at 4.4490 GHz. We set the LO frequency to the frequency of steepest slope to maximize the sensitivity. The FWHM of the averaged signal is 6.4 MHz, which is slightly larger than the average FWHM of individual spectra [red solid line in Fig.~\ref{3000}(g)].

\begin{figure}
\includegraphics[width=\linewidth]{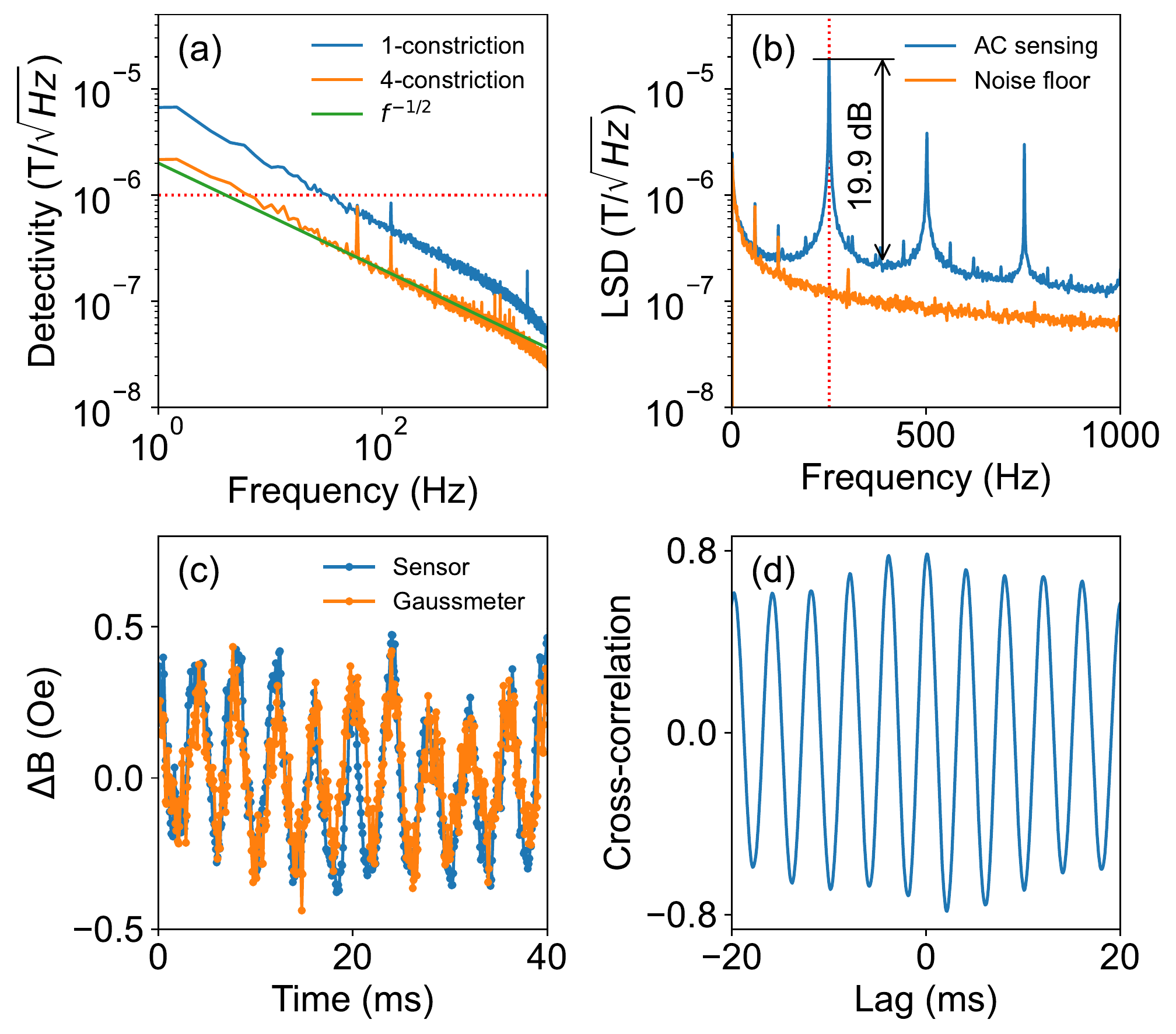}
\caption{\label{AC} (a) Detectivity of 1- and 4-constriction SHNO sensors, showing a comparison with $f^{-1/2}$ slope. The dotted line indicates $1\ \mu$T$/\sqrt{\mbox{Hz}}$. (b) Linear spectral density of a 251 Hz modulation magnetic field with rms 0.153 Oe, measured by the 4-constriction sensor. The noise floor from (a) is also plotted for comparison. The dotted line indicates 251 Hz. (c) Measurements of the same magnetic field in the time domain, showing comparison with a commercial Gaussmeter (Lakeshore 475) probe (Lakeshore HMNT-4E04-VR) (d) Normalized cross-correlation of two signals in (c).}
\end{figure}

For the 4-constriction device, substituting the values $\partial V/\partial f$ extracted from Fig.~\ref{3000}(c) and $\partial f/\partial H$ from Fig.~\ref{Auto}(c)  into Eqn.~(1), we obtain the sensitivity  $S=2.04$ V/Oe. By measuring the linear spectral density (LSD) of the noise, the detectivity (i.e. field equivalent noise)\cite{reig2013giant} of the 4-constriction device can be obtained using
\begin{equation}
D(\mbox{Oe}/\sqrt{\mbox{Hz}})=\frac{\mbox{LSD}(\mbox{V}/\sqrt{\mbox{Hz}})}{S(\mbox{V/Oe})}. 
\end{equation}
We measure the noise LSD by taking the square root of 100-iteration averaged periodogram of the voltage signal. The periodogram is an estimate of the power spectral density, and is calculated with a Fourier transform. 
 Substituting in the sensitivity value, we plotted the detectivity of the 4-constriction device in [Fig.~\ref{AC}(a)]. Similarly, we characterize the detectivity of the 1-constriction device for comparison. As expected, the 1-constriction device has a larger detectivity compared to the 4-constriction one since the latter has a more stable frequency. In both cases the detectivity decreases with frequency, which is common to many electronic  sensors.\cite{davies2021magnetoresistive} For an externally applied AC magnetic field of 100 Hz, the detectivity of the 4-constriction device is $0.21\ \mu$T$/\sqrt{\mbox{Hz}}$, while the detectivity of 1-constriction device is $0.53\ \mu$T$/\sqrt{\mbox{Hz}}$. The slope of both devices are close to $1/\sqrt{f}$, i.e.~$1/f$ scaling in power spectral density (PSD), which indicates that our SHNO sensor is dominated by flicker noise.\cite{ramsden2011hall}

\begin{figure}
\includegraphics[width=\linewidth]{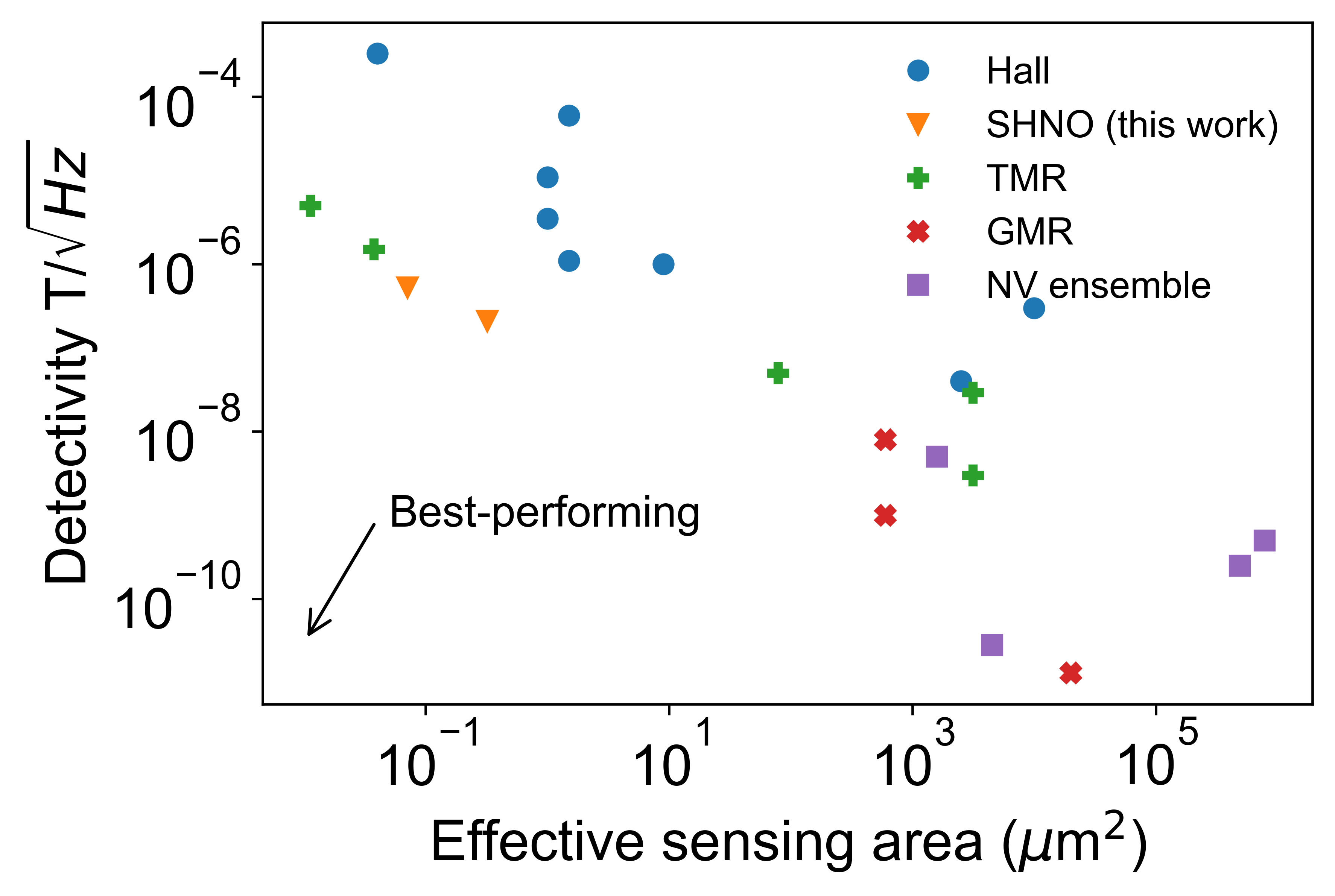}
\caption{\label{Detectivity} Detectivity at 100 Hz versus effective sensing area of different room temperature magnetic field sensors, including Hall sensor,\cite{sandhu2004nano,shiogai2019low,joo2016large,mihajlovic2007inas,dauber2015ultra,schaefer2020magnetic} tunneling magnetoresistance (TMR) sensor,\cite{monteblanco2021constant,davies2021magnetoresistive,leitao2014nanoscale} giant magnetoresistance (GMR) sensor\cite{cubells2016integration,reig2013giant,kikitsu2022magnetic} and NV ensemble sensor.\cite{webb2019nanotesla,chen2022enhancing,zheng2019zero,chatzidrosos2017miniature}}
\end{figure}

To test our 4-constriction device in measuring an external AC magnetic field, we applied a 251 Hz AC modulating field with rms of 0.153 Oe parallel to the bias magnetic field, and compare the result with a commercial gaussmeter (Lakeshore 475) probe (Lakeshore HMNT-4E04-VR). We obtain signal in the time domain [Fig.~\ref{AC}(c)], and extract the frequency domain information by taking the square root of 100-iteration averaged periodogram of the voltage signal [Fig.~\ref{AC}(b)]. In the frequency domain, peaks at 251 Hz and its harmonics are visible at a signal level that is 19.9 dB above background. This shows the ability of our SHNO sensor to sense the correct frequency of AC magnetic field. The noise floor from the detectivity measurement is plotted in Fig.~\ref{AC}(b) for comparison. The baseline of the 251 Hz sensing signal is higher than the noise floor for frequency > 50 Hz. This may be due to added noise associated with the AC field generation. In the time domain, the signal from SHNO sensor shows temporal similarity with the gaussmeter response as seen from the normalized cross-correlation of the two signals [Fig.~\ref{AC}(d)]. The maximum cross-correlation is 0.785, indicating a good match. The normalized cross-correlation also reveals the periodicity of both signals, indicating both signals successfully follow the AC magnetic field. 

To visualize the performance of our SHNO sensors with respect to other room temperature analog magnetic field sensors, we plot their detectivity with respect to their effective sensing area (total active area of the sensing elements) in Fig.~\ref{Detectivity}.  We exclude single diamond nitrogen-vacancy (NV) center sensors in this comparison because they are not operated as analog sensors, however, NV center ensemble sensors typically use analog detectors and are appropriate for comparison.  In general, there is a phenomenological trade-off between detectivity and the effective area of the sensing element. The best-performing sensors, which combine small sensing area and low detectivity, will be at the bottom left corner of the plot. It is noteworthy that our SHNO sensor has a lower detectivity compared to other sensors of similar sub-micron effective sensing area and appear at the frontier of detectivity closest to the desirable corner among a wide class of sensors. We envision a further reduction on the detectivity by making more advanced SHNO sensors that have a magnetic tunneling junction (MTJ) at the constriction area to provide a larger magnetoresistive voltage output.

In conclusion, we have demonstrated SHNO devices as nanoscale magnetic field sensors for low frequency magnetic field sensing. We show a detectivity for a 100 Hz AC magnetic field as low as $0.21\ \mu$T$/\sqrt{\mbox{Hz}}$ with a effective sensing area of 0.32 $\mu$m$^2$. This nanoscale sensing area enables application of SHNO sensors as local sensor, such as in scanning probe magnetometry.

\vspace{2ex}
This work was primarily supported by the Cornell Center for Materials Research with funding from the NSF MRSEC program (DMR-1719875).  Device fabrication and some preliminary work was supported by the NSF (ECCS-1708016). This work was performed in part at Cornell NanoScale Facility, an NNCI member supported by NSF Grant NNCI-2025233.

\bibliographystyle{apsrev4-2} 
\bibliography{aipsamp}

\end{document}